\documentclass[11pt]{article}
\usepackage{fleqn,cospar}

\usepackage{graphicx}

\title{
ON A PROBABLE CONTRIBUTION \\
OF THE EINSTEIN--PODOLSKY--ROSEN EFFECT \\
TO SOLVING THE PROBLEM OF HOMOGENEITY \\
OF THE UNIVERSE
}

\author{Yu.V. Dumin\address{IZMIRAN, Russian Academy of Sciences,
                            Troitsk, Moscow region, 142190 Russia}}

\begin{document}

\maketitle

\begin{abstract}

Evading formation of the domain walls in
cosmological phase transitions is one of key problems to be solved
for explanation of the observed large-scale homogeneity of the Universe.
The previous attempts to get around this obstacle
led to imposing severe observational constraints on
the parameters of the fields involved.
Aim of the present report is to show that yet another way
to overcome the above problem is accounting for
Einstein--Podolsky--Rosen (EPR) effect.
Namely, if the scalar (Higgs) field was presented by
a single coherent quantum state at the initial instant of time,
then its reduction during a phase transition at some later instant
should be correlated even at distances
exceeding the local cosmological horizon.
Consideration of a simplest one-dimensional cosmological model
with $Z_2$ Higgs field demonstrates that
EPR correlations can substantially reduce the probability of
spontaneous creation of the domain walls,
and such effect seems to be even more promising
in the cases of two- and three-dimensional geometry.

\end{abstract}

\section*{INTRODUCTION}

The problem of domain-wall formation during spontaneous breaking of
discrete symmetry was emphasized for the first time by
Zel'dovich et al. (1975) just when the role of Higgs fields
in cosmology began to be recognized. This problem is associated with
the fact that the observed region of the Universe contains
a large number of the domains that were causally-disconnected
at the instant of phase transition, and the stable vacuum states
of such domains after the symmetry breaking, in general, should differ
from each other. As a result, a network of domain walls,
involving considerable energy density, will be formed.
But, on the other hand, the presence of such domain walls is
incompatible with the observed large-scale homogeneity of the Universe.

The previously-undertaken attempts to resolve the above problem
were based on imposing severe constraints on the parameters or
modifying the field theories involved
(see, for example, review by Larsson et al., 1997).

Aim of the present report is to show that, in principle,
the domain-wall problem may be resolved by a natural way
if we take into account the fact that Higgs condensate represents
a single coherent quantum state,%
\footnote{Just the assumption of coherent
(sometimes also called ``classical'') character of the Higgs field
is a key item in the mechanism of elementary-particle mass generation.}
which should experience
Einstein--Podolsky--Rosen (EPR) correlations (Einstein et al., 1935)
even in causally-disconnected regions
during its reduction to the state of broken symmetry.

Such point of view seems to be especially attractive due to
the recent experimental achievements, such as
(a)~the quantum-optical experiments that confirmed a presence of
EPR correlations of the single photon states at
considerable distances ($\sim$10 km), and
(b)~the studies of Bose--Einstein condensate of ultracooled gases,
which demonstrated that all predictions of
the ``orthodox'' quantum mechanics are valid for a coherent quantum state
involving even a macroscopic amount of substance.

So, if we believe that EPR correlations really take place in
Higgs condensate, then it should be expected that
the probabilities of various realizations of
the Higgs-field configurations after the symmetry breaking
will be distributed by Gibbs law. As a result,
the high energy concentrated in the domain walls
(and, therefore, contradicting the astronomical observations)
turns out to be just the reason why probability of the respective
configurations is strongly suppressed (Dumin, 2000).

The basic question arising here concerns the efficiency of
such suppression for a particular set of the field parameters.
The same problem can be reformulated by
an opposite way:
What parameters should the Higgs field
(and its phase transition) have for the probability of
domain-wall formation to be substantially reduced?
We shall try to give a quantitative answer to this question
in the next section.

\section*{THE MODEL OF PHASE TRANSITION ALLOWING FOR EPR CORRELATIONS}

Let us consider the simplest one-dimensional cosmological model
with metric
\begin{equation}
ds^2 = \: dt^2 - a^2(t) \, dx^2 .
\label{init_metric}
\end{equation}
By introducing the conformal time $ \eta = \int dt / a(t) $,
Eq.~\ref{init_metric} can be reduced to the form
$ \, ds^2 = \, a^2(t) \: \{ d{\eta}^2  - dx^2 \}; \, $
so that the light rays
($ ds^2 = \, 0 $)
are described as
$ \, x = \, \pm \, {\eta} \, + \, {\rm const}$~(e.g. Misner, 1969).

\begin{figure}
\centerline{
\includegraphics[width=8cm]{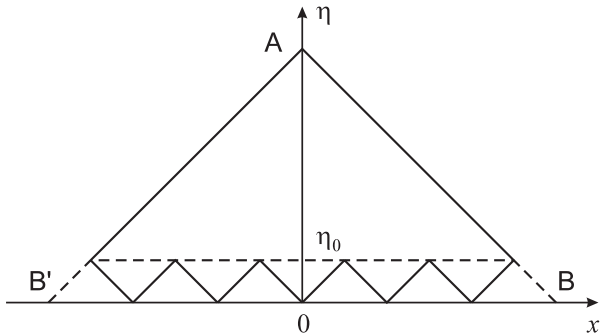}
}
\caption{Conformal diagram of the space--time under consideration.}
\end{figure}

Let $ {\eta}_{\scriptscriptstyle 0} $ be the conformal time
corresponding to a phase transition.
Then, the region of the Universe observed at the instant $\eta$
will contain $N$ domains that were causally-disconnected during
the phase transition (see Figure~1):
\begin{equation}
N = \,
  ( {\eta} - {\eta}_{\scriptscriptstyle 0} ) /
    {\eta}_{\scriptscriptstyle 0} \: \approx \:
  {\eta} \, / {\eta}_{\scriptscriptstyle 0} \: .
\end{equation}

The Higgs field $\varphi$, possessing $Z_2$ symmetry group,
after reduction to the state of broken symmetry can be
in one of two stable vacua
($ +{\varphi}_{\scriptscriptstyle 0} \, $ or
$ \, -{\varphi}_{\scriptscriptstyle 0} $)
in each of the above-mentioned domains.
The energy of a ``wall'' between two domains with different vacua
will be denoted by $E$ (its expression in terms of
the Lagrangian parameters can be found, for example,
in paper by Zel'dovich et al, 1975).
Besides, for the sake of self-consistency,
the periodic boundary conditions will be imposed at
the opposite sides of the observable region (i.e.,
the boundary $\tt AB$ will be identified with $\tt AB'$);
so that the possible total number of the domain walls
is always even.

So, under assumptions formulated above,
the probability of realization of the Higgs-field configuration
involving $2k$ domain walls is equal to
\begin{equation}
P^{2k}_{\scriptscriptstyle N} \, = \:
  2 \, A_{\scriptscriptstyle N} \,
  \frac{N!}{(2k)! \; (N \! - \! 2k)!} \: e^{-2kE/T} ,
\end{equation}
where $T$ is the characteristic temperature of phase transition,
and $A_{\scriptscriptstyle N}$ is the normalization factor defined as
\begin{equation}
A^{\scriptscriptstyle -1}_{\scriptscriptstyle N} \, = \:
  \sum^{\scriptscriptstyle [N/2]}_{k=0} \:
  \frac{2 \: N!}{(2k)! \; (N \! - \! 2k)!} \: e^{-2kE/T} ,
\label{norm_factor}
\end{equation}
where square brackets in the upper limit of the sum denote
the integer part of a number.

In principle, the problem of finding the normalization factor
given by Eq.~\ref{norm_factor} can be formally reduced to
determination of the statistical sum for one-dimensional Ising model
(e.g. Huang, 1963 and Isihara, 1971), which has an exact solution.
Nevertheless, to get the most physical clarity, we shall not use here
this general approach and restrict our consideration only by
limiting cases of high and low temperatures
(in comparison with the energy of a domain wall).
As will be seen later, the respective asymptotic formulas can be
reasonably fitted to each other.

\subsection*{The Case of High Temperature}

In the case of high temperature,
when neighboring terms of the sum differ from each other only slightly,
summation can be approximately conducted over all numbers
of the domain walls (both odd and even) and finally gives
\begin{equation}
A^{\scriptscriptstyle -1}_{\scriptscriptstyle N} \, \approx \:
  {\left( 1 + e^{-E/T} \right)}^{\scriptscriptstyle N} ;
\end{equation}
so that
\begin{equation}
P^{2k}_{\scriptscriptstyle N} \, \approx \:
  \frac{2 \: N!}{(2k)! \; (N \! - \! 2k)!} \;
  \frac{e^{-2kE/T}}%
  {{\left( 1 + e^{-E/T} \right)}^{\scriptscriptstyle N}} \; .
\end{equation}

In particular, a probability of the Higgs-field configuration without
any domain walls (which is just the case actually observed in the Universe)
equals
\begin{equation}
P^{0}_{\scriptscriptstyle N} \, \approx \:
  \frac{2}{{\left( 1 + e^{-E/T} \right)}^{\scriptscriptstyle N}} \; .
\end{equation}

It is interesting to find restriction on the domain-wall energy $E$
and phase-transition temperature $T$ for
the probability of absence of the domain walls
to be greater than or equal to some specified number~$p$
($ 0 < p < 1 $; for example, $ \, p = 1/2 $):
\begin{equation}
E/T \, \ge \,
  \ln \frac{1}{(2/p)^{\scriptscriptstyle 1/N} - 1} \, \approx \,
  \ln N - \ln \ln \, (2/p) \: \approx \:
  \ln N \, .
\label{E-T_high_temp}
\end{equation}

\subsection*{The Case of Low Temperature}

As regards the opposite case of low temperature,
when the main contribution to the sum in Eq.~(\ref{norm_factor})
is produced by a few first terms,
we can approximately take into account only two of them; so that
\begin{equation}
A^{\scriptscriptstyle -1}_{\scriptscriptstyle N} \, \approx \:
  2 \: \left\{ 1 + \frac{1}{2} \, N^{2} \, e^{-2E/T} \right\} ;
\end{equation}
and the respective probability of the Higgs-field configuration
involving $2k$ domain walls will be
\begin{equation}
P^{2k}_{\scriptscriptstyle N} \, \approx \:
  \left\{ 1 - \frac{1}{2} \, N^{2} \, e^{-2E/T} \right\}
  \frac{N!}{(2k)! \; (N \! - \! 2k)!} \; e^{-2kE/T} .
\end{equation}

Then, the probability that domain walls are absent at all equals
\begin{equation}
P^{0}_{\scriptscriptstyle N} \, \approx \:
  1 - \frac{1}{2} \, N^{2} \, e^{-2E/T} ,
\end{equation}
and it will be greater than or equal to the specified number~$p$ if
\begin{equation}
E/T \, \ge \,
  \ln N - \frac{1}{2}  \, \ln \, ( 2 \, (1 \!\! - \! p) ) \: \approx \:
  \ln N \, ,
\label{E-T_low_temp}
\end{equation}
i.e., to a first approximation, it is again determined by
the logarithmic function of $N$.

\bigskip

So, as follows from Eqs.~\ref{E-T_high_temp}
and~\ref{E-T_low_temp},
the ratio $ E/T $ differs from unity only by $ \ln N $.
Consequently, the required temperature of the phase transition is
of the same order of magnitude as the domain-wall energy
even at large value of~$N$.
Therefore, the required parameters
may be satisfied in some particular kind of the field theory.

\section*{DISCUSSION}

There is no doubt that a considerably more sophisticated analysis
must be carried out to draw unambiguous conclusion on the role of
EPR correlations in the phase transitions of Higgs fields.
Nevertheless, our consideration,
based on the simplest one-dimensional model,
shows that such possibility could be quite promising.

Moreover, the situation seems to be even better in the cases of
two- and three-dimensional geometry.
A well-known property of Ising models in two and three dimensions
(as distinct from the one-dimensional case) is a presence of
second-order phase transition
at some temperature on the order of the domain-wall energy.
As a result, one can expect that a restriction on $ E/T $,
similar to Eqs.~\ref{E-T_high_temp} and~\ref{E-T_low_temp},
will not contain dependence on $N$ at all.%
\footnote{To avoid misunderstanding, it should be emphasized that
the Ising model, which can be introduced to study
the long-range correlations during the phase transition,
is only an auxiliary mathematical construction.
So, a phase transition in the Ising model is in no way identical to
the real physical phase transition of the respective Higgs field.
(Particularly, the Ising model can possess a phase transition of
only the second order or experience no phase transition at all;
whereas the physical phase transition of the Higgs field can be
both of the first or second order,
depending on the particular Lagrangian.)}
This problem is still to be studied more carefully.

\end{document}